\documentclass[alpha-refs]{wiley-article}

\usepackage{siunitx}

\usepackage{moreverb}

\usepackage{graphicx}
\usepackage{amsmath, amssymb} 
\usepackage{array}
\usepackage{verbatim} 
\usepackage{scrextend}
\usepackage{footmisc}

\usetikzlibrary{fit}					
\usetikzlibrary{backgrounds}	
\usetikzlibrary{shapes,decorations}
\usepackage{wrapfig} 
\usepackage{multirow}

\papertype{Original Article}
\paperfield{Journal Section}

\title{Associating Growth in Infancy and Cognitive Performance in Early Childhood:  
A functional data analysis approach}

\abbrevs{FDA, functional data analysis, FPCA, functional principal component analysis.}

\author[1]{Pantelis Z. Hadjipantelis}
\author[1]{Kyunghee Han}
\author[1]{Jane-Ling Wang}
\author[2]{Seungmi Yang}
\author[2]{Michael S. Kramer}
\author[3, 4]{Richard M. Martin}
\author[5]{Emily Oken}
\author[1]{Hans-Georg M\"uller}

\affil[1]{Department of Statistics, University of California, Davis, California, USA}
\affil[2]{Departments of Pediatrics and of Epidemiology, Biostatistics and Occupational Health, McGill University, Montreal, Canada}
\affil[3]{Bristol Medical School, Population Health Sciences, University of Bristol, UK}
\affil[4]{National Institute for Health Research Bristol Biomedical Research Centre, University Hospitals Bristol NHS Foundation Trust and the University of Bristol, Bristol, UK}
\affil[5]{Obesity Prevention Program, Department of Population Medicine, Harvard Medical School and the Harvard Pilgrim Health Care Institute, Boston, Massachusetts, USA}

\corraddress{Pantelis Z. Hadjipantelis}
\corremail{pantelis@ucdavis.edu}

\fundinginfo{This study was supported by the Human Birth, Growth and Development Knowledge Integration (HBGD\emph{ki}) project of the Bill and Melinda Gates Foundation.}

\runningauthor{Hadjipantelis et al.}

\begin{document}

\maketitle

\begin{abstract}
\small
Physical growth traits can be naturally represented by continuous functions. In a large dataset of infancy growth patterns, we develop {a practical approach to infer statistical associations between} growth-trajectories and IQ performance in early childhood. The main objective of this study is to show how to assess physical growth curves and {detect} if particular infancy growth patterns are associated with differences in IQ (Full-scale WASI scores) in later ages using a semi-parametric functional response model. Additionally, we investigate the association between different growth measurements in terms of their cross-correlation with each other, their correlation with later IQ, as well as their { time-varying dynamics.} 
This analysis framework can easily incorporate or learn population information in a non-parametric way, rendering the existence of prior population charts partially redundant.

\keywords{Functional Principal Component Analysis, Functional Concurrent Regression, Early-childhood IQ}
\end{abstract}

\section{Introduction} \label{Introd}

There is increasing awareness that early childhood growth patterns may be associated with adult cognitive performance and other traits \citep{smithers2013impact,wright2015head}. Coupled with the increasingly larger  and more carefully annotated data that are becoming available, this has lead to increased interest in the analysis of human growth curves.
Growth curves are commonly modelled as a (small) number of discrete time-correlated measurements; a functional data analysis (FDA) framework is rarely employed \citep{horvath2012inference} despite the fact that FDA approaches provide a flexible non-parametric approach to modelling growth patterns. While certain biological applications (e.g. \cite{chen2012conditional}) have been presented in the past, only recently  have large human population samples been used in FDA modelling \citep{zhang2015regression}. Even then the analysis is mostly macroscopic and does not examine the association of specific growth patterns with external variables nor the interplay between functional growth trajectories with a response variable. Here we illustrate specifically the application of FDA in investigating the relation between infant growth curves of head circumference and body length while considering  early childhood IQ as a response variable.

One of the basic findings in most studies examining head circumference growth and IQ performance is their strong positive association. 
In \cite{gale2006influence} the highest IQ scores (at 4 years) were observed for children: ``\textit{whose head circumference had grown large prenatally and whose head circumference during infancy had grown larger than expected, given its size at birth}".  
Similarly, children with consistently small heads or no catch-up growth had lower IQ (and/or neurocognitive disorders) in later ages \citep{wright2015head,fattal2009growth}. 
It is therefore an open question how to accurately associate IQ and infancy growth patterns.
To provide new information that helps the better understanding of this association we utilise two core FDA methodologies:
functional principal component analysis (FPCA) \citep{Yao05} and functional concurrent regression (FCR) \citep{csenturk2011varying}. 
FPCA is able to provide parsimonious and informative characterisations of function-valued data like growth curve trajectories, within a nonparametric statistical framework. Similarly, FCR allows us to quantify the time-domains where the influence of a particular variable on an outcome is strong. In particular we aim to identify the time where the association of infancy growth curve attains its maximal correlation with early childhood IQ and if particular growth patterns in body length and head circumference are associated with early childhood IQ. 

\section{Growth curve modelling by functional data analysis}\label{s:methods}
 
In this work, growth curve analysis is conducted through a functional principal component analysis of the sparsely sampled growth curves.  Many previous studies of pediatric growth are limited to multivariate techniques that do not fully utilise assumptions about the underlying dynamics of growth curves despite the natural representation of growth curves as functional data \citep{gale2006influence,fattal2009growth,smithers2013impact,wright2015head}. 
Functional data assume that the unit of analysis is bounded continuous curve, rather than a time-indexed vector of readings. 
Early work of Gasser et al. has shown that the functional analysis approach presents a viable alternative to standard parametric techniques \citep{Gasser1984}, where kernel estimators offered relevant insights when analysing irregularly sampled data of human growth patterns \citep{kneip1992statistical,ramsay1995comparison}. More recently, Park and Ahn have used FDA methodology to create efficient multivariate clusters of growth patterns \citep{park2016clustering}. In a similar manner \citep{zhang2015regression} screened the entire growth path to detect potentially problematic growth patterns and \cite{leroux2017dynamic} showcase how to functional concurrent regression can be applied in the context of dynamic prediction of growth paths. \cite{wang2016functional} provide a detailed review of FDA. 
  
Functional variables will be denoted by $X(\cdot)$ (or $X$ for brevity) and scalar variables will be denoted with $Z$. Using the original notion of a stochastic process $X$ \citep{Castro86}, the optimal $K$-dimensional linear approximation in the sense of minimizing the variance of the residuals is: 
\begin{align} 
  {X}_i(t) {~\approx~} \mu_X(t)+ \sum_{k=1}^{K} \xi_{i,k} \phi_k(t),	\label{CastroEq}  
\end{align}
where the functional mean is defined as $ \mu_{X}(t) = E\{X(t)\}$,
$\xi_{i,k}$ are the $k$-th principal component scores and $\phi_k$ ($k \geq 1$) are the eigenfunctions that form {  an orthonormal basis of a space of square-integrable functions}; the index $i$ refers to the $i$-th subject. As in the case of standard multivariate PCA, the fraction of the sample variance explained is maximised along the modes of variation represented by the eigenfunctions $\phi_k$.  

Utilising FPCA, we identify the principal modes of amplitude variation in the sample $X$ and use those modes as a basis to project our data to a finite-dimensional subspace by imposing a finite truncation point on the number of basis terms. Specifically, we define the  auto-covariance function $C_X$ as: 
\begin{equation}
C_X(s,t) = E\{\left(X_i(s)-\mu_X(s)\right)\left(X_i(t^{})-\mu_X(t^{})\right)\}. \label{AutoCovOper}  
\end{equation}
By Mercer's theorem\citep{Mercer09}, the spectral decomposition of the symmetric amplitude auto-covariance function ${C}_X$ can be written as:
\begin{align}
 {C}_X(s,t^{}) = \sum_{{k}=1}^\infty \lambda_{k} {\phi}_{k}(s)  {\phi}_{k}(t^{}),
\end{align}
where the eigenfunctions ${\phi_k}$ are treated as the FPCA-generated continuous modes of variation. Additionally, the eigenvalues $\lambda_{k}$ allow the determination of the total percentage of variation exhibited by the sample along the $k$-th principal component and indicate whether the examined component is relevant for further analysis. The choice of the number of components can be based on a fraction-of-variance-explained criterion (e.g. 95\%). Directly examining the data themselves supports this selection, as non-included higher order components have very low variance (Table \ref{FVEtable}). 
Having fixed $K$ as the number of eigenfunctions $\phi$ (i.e. functional principal components - FPCs) to retain, we use these eigenfunctions to compute $\xi_{i,k}$, the amplitude projection scores associated with the $i$-th sample and the $k$-th component (Eq. \ref{PCscoresAmp}) as: 
\begin{align}
  {\xi}_{i,k} = \int\{X_i(t) - \mu_{X}(t)\} {\phi_k(t)} dt,\label{PCscoresAmp}
\end{align}
where a suitable numerical approximation to the integral is used for practical analysis for $k \in \{1, \dots, K\}$. For cases with relatively sparse data, we use a probabilistic approximation based on conditioning upon the observed data \citep{Yao05}. In that case the amplitude projection scores associated with the $i$-th sample and the corresponding $k$-th FPCs (Eq. \ref{PCscoresAmp}) are:
\begin{align}
  \hat{\xi}_{i,k} = E[\xi_{i,k}|X_i ]  =  {\lambda}_k \phi_{i,k}^T {C}_{X_i}^{-1}(X_i- \hat{\mu}_i). \label{compPACE}
\end{align}
The elements defined in Eq. \ref{compPACE} are all unknown and are estimated nonparametrically directly from the data. The exact local least square estimators and formulas with the observed data as inputs are presented in the Appendix, Sect. \ref{app1}. 

In a similar manner as for $C_X$, we define the cross-covariance function between processes $X_1$ and $X_2$ \citep{Yang2011} as:
 \begin{align}
   C_{X_1,X_2}(s,t) = E\{\left({X_1}_i(s)-\mu_{X_1}(s)\right)\left({X_2}_j(t)-\mu_{X_2}(t)\right)\}. \label{CrossCovOper}
 \end{align}
%
Note that the cross-covariance matrix is asymmetric (i.e. $C_{X_1,X_2}[s,t] \neq ( C_{X_1,X_2}[s,t])^T$) and implies the cross-correlation surface between two functional variables $X_1$ and $X_2$:
\begin{align}
  R_{X_1,X_2} = D_{X_1}^{\frac{1}{2}} C_{X_1,X_2} D_{X_2}^{\frac{1}{2}},
\end{align}
where $D_{X_i}$ is the diagonal matrix holding the inverse of the diagonal values of the autocovariance $C_{X_i}$. Using the  above definitions of auto- and cross- covariances (Eq. \ref{AutoCovOper} \&  \ref{CrossCovOper}), we can derive a functional concurrent regression (FCR) model \citep{csenturk2010functional,csenturk2011varying} that allows investigating the time-varying influence a particular covariate on the outcome of interest. This is done by pooling information from all subjects and building smooth trajectories across time. A functional concurrent multiple regression model for zero-centred dependent variables $X^C(t)$ is given by:
\begin{align}
E\{ X^C(t)|X_1(t), \dots, X_p(t), Z_1, \dots, Z_q\} = \sum_{r=1}^p \alpha_r(t) X_r(t) +  \sum_{g=1}^{{q}} \beta_g(t) Z_g,
\end{align}
where the varying coefficient functions of interest are obtained as:
\begin{align}
\left( \begin{array}{c}
\alpha_1(t)\\
\vdots\\
\alpha_p(t)\\
\beta_1(t)\\
\vdots\\
\beta_q(t)
\end{array}\right) 
= 
\left( \begin{array}{cccccc}
      C_{X_1}(t,t) & \dots & C_{X_1,X_{p}}(t,t) & C_{X_1,Z_1}(t) & \dots & C_{X_1,Z_{q}}(t)\\
      \vdots & \ddots & \vdots & \vdots & \ddots & \vdots\\
      C_{X_1,X_{p}}(t,t) & \dots & C_{X_{p}}(t,t) & C_{X_{p},Z_1}(t) & \dots & C_{X_{p},Z_{q}}(t)\\
      C_{X_1,Z_1}(t) & \dots & C_{X_{p},Z_1}(t) & C_{Z_1}  & \dots & C_{Z_1,Z_{q}}\\
      \vdots & \ddots & \vdots & \vdots & \ddots & \vdots\\
      C_{X_1,Z_{q}}(t) & \dots & C_{X_{p},Z_{q}}(t) & C_{Z_1,Z_{q}} & \dots & C_{Z_{q}} \\
      \end{array} \right)^{-1}\ \left( \begin{array}{c}
                                      C_{{{X^C}},X_1}(t,t) \\
                                      \vdots\\
                                      C_{{{X^C}},X_{p}}(t,t) \\
                                      C_{{{X^C}},Z_1}(t) \\
                                      \vdots\\
                                      C_{{{X^C}},Z_{q}}(t) \\
                                      \end{array} \right). 
\label{CrossCovOperReg}
\end{align} 
As with Eq. \ref{compPACE}, the estimators used by Eq. \ref{CrossCovOperReg} are estimated nonparametrically directly from the data. Their computation is outlined in the Appendix, Sect. \ref{app2}. When implementing the functional concurrent regression described above we use the centred and scaled version of the longitudinal covariates $X$. We achieve the centring and scaling by employing a procedure similar to that of \cite{chiou2014multivariate} which utilises the sample's smoothed mean $\mu_X(t)$ and auto-covariance function $C_X(s,t)$. The exact estimation procedure for the FCR is given in \cite{csenturk2011varying}.

\section{Sample Pre-processing and Preliminary FPCA investigation}

\begin{table}[!ht]
\begin{center}
\caption{The list of variables currently examined.}\label{TableVars}
\begin{threeparttable}
\begin{tabular}{llcc}
\headrow
\thead{Variable type} & \thead{Variable name} & \textbf{Unit} & \textbf{Abbreviation}\\
\hiderowcolors
\multirow{2}{*}{Time-varying $X(t)$}	&	Head circumference	&	cm	&	$X^H(t)$\\
	&	Length (or Height)	&	cm	&	$X^L(t)$\\
\hline
\multirow{7}{*}{Time-invariant $Z$}	&	Full-scale IQ at 6.5 years	&	IQ	&	$IQ$\\
	&	Birth-weight	&	 kg	&	 $BW$\\
	&	Sex	&	  binary	&	  $Sex$\\
	&	Maternal (Paternal) Education level	&	 ordinal	&	 $ME$ ($PE$)\\
	&	Maternal (Paternal) Age-at-Birth	&	years	&	 $MAB$ ($PAB$)\\
	&	Duration of exclusive breast-feeding	&	ordinal	&	 $EBF$\\
	&	Hospital	&	nominal	&	 $Hospital$\\
\hline  
\end{tabular}

\end{threeparttable}
\end{center}
\end{table}

To the best of our knowledge, this report is the first application of an FDA framework targeting the developmental questions outlined in Sect. \ref{Introd}. We demonstrate the application of these concepts with the Promotion of Breastfeeding Intervention Trial (PROBIT) \citep{kramer2001promotion,kramer2003infant, kramer2004optimal} dataset.
PROBIT is a cluster-randomised controlled trial of a breastfeeding promotion intervention modelled on the WHO/UNICEF Baby-Friendly Hospital Initiative in the Republic of Belarus. 
17046 healthy term infants who weighed $\geq 2500$ g were recruited from 31 maternity hospitals and affiliated polyclinics at birth and were followed up at 1, 2, 3, 6, 9, and 12 months as well as at age 6.5 years. At 6.5 years study paediatricians measured cognitive ability using the Wechsler Abbreviated Scale of Intelligence (WASI). 
The variables used  in this study are shown in Table \ref{TableVars}. Head circumference and body length are abbreviated as  $X^H(t)$ and  $X^L(t)$ respectively, and are recorded at variable times $t$. $IQ$ corresponds to the full-scale WASI score at 6.5 years of age.  The weight at birth is abbreviated by $BW$ and the sex of the child as $Sex$. Maternal and paternal education levels are ordinal variables and abbreviated as $ME$ \& $PE$ respectively; the available levels are (completed) less than secondary education, common secondary education, some post-secondary education and tertiary education.  
Duration of exclusive breast feeding is also an a ordinal variable ($EBF$); the available levels are: up to three months, three to six months, six months or more.  
Information used also includes: maternal and paternal age-at-birth ($MAB$ \& $PAB$) and maternal smoking during pregnancy ($MS$).
Finally we also used the information regarding the hospital in which the child was born ($Hospital$). This information is crucial as it indirectly encapsulates socio-economic information both for the parents as well as for the environment the child grew up.

The vast majority of the curves analysed 7 readings per child (93.5\%), while a smaller proportion of the sample analysed has 6 (5.5\%). 
Although the follow-up examinations were scheduled at specific months as described above, the resulting design plot (Appendix, Sect. \ref{app3}, Fig. \ref{DesignPlot}) shows that the actual time measurements were not followed exactly. This non-adherence to the rigid design points has advantages for the analysis because it allows a more uniform coverage across the time-domain examined. Details on the construction of design plots, as proposed by \cite{Yao05}, are given in the Appendix, Sect. \ref{app3}. 
716 children who missed two or more scheduled visits in the first year were excluded from further analysis because missing values might not have occurred at random but could potentially indicate some problems with the data acquisition for these children. 
In addition, we excluded children with potential data entry errors (0.2\%) 
(e.g. when a child's three-month visit date was recorded earlier than its two-month visit date or  children growing 8.5 cm in head circumference within a single month). 
However, we do not impose monotonicity constraints in the longitudinal variables examined. While quite possibly monotonic for healthy children, we treat deviations from monotonicity as measurement errors. 
Our final sample included 12,809 children, after also excluding children that had missing demographic information (e.g. maternal education).

\begin{figure}[!ht]
 \centering
    \includegraphics[width=0.8\textwidth]{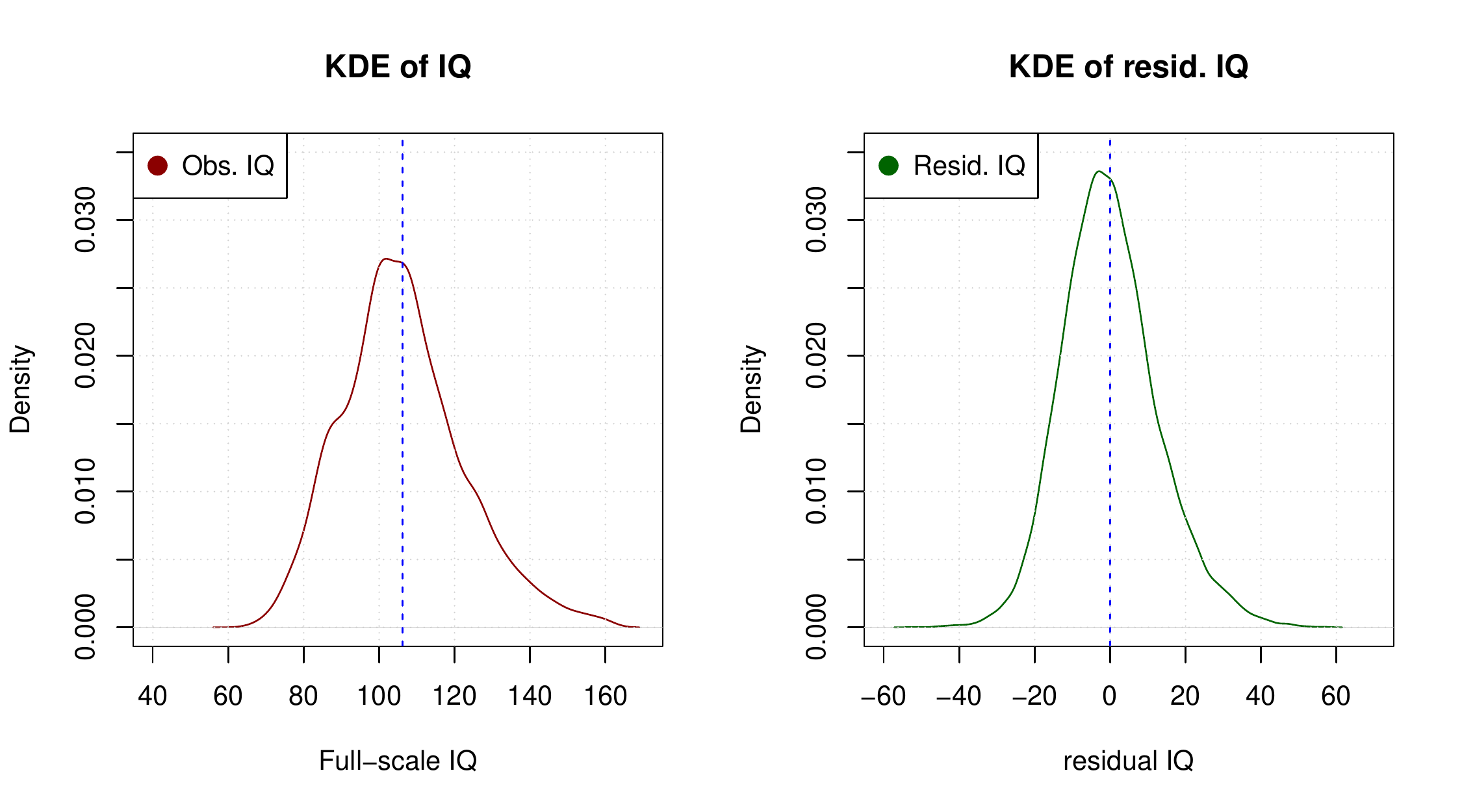} 
  \caption{Left hand-side plot: The observed WASI score kernel density estimate,  $IQ_{full}$. Right hand-side plot: The distributional kernel density estimate of the residual IQ, $IQ_{res}$. $IQ_{full}$ corresponds to the observed WASI score measure at 6.5 years of age. $IQ_{res}$ corresponds the IQ residuals variates that remain after marginalising out the influence of covariates fixed at birth (e.g. maternal and paternal educational characteristics, sex of the child, hospital of birth, etc.) 
}\label{IQs}
\end{figure}

Computing the mean of a functional variable $\mu_X$ is important as a first step, where we aim to estimate smooth, twice differentiable functions. The smoothness assumption is essential so that the discretely observed sample data can be viewed as functional \citep{horvath2012inference}. To conduct FPCA with conditional expectation (PACE) \citep{Yao05}, we do not smooth the data of individuals but rather apply smoothing to aggregated data. In line with \cite{Chiou03}, we use a locally weighted least squares smoother, $S_{1D}$, in order to fit a local linear polynomial to the aggregated data of each longitudinal variable. That estimate is then treated as the sample's smooth mean. 
Conceptually, the value of the smoothed curve at a point $t$ in a smoothing window $[t - b, t + b]$ is calculated as the intercept of the local regression line among the data falling into the smoothing window, when predictors are centred at the midpoint of the window. The kernel function $K$ used is the Gaussian kernel function $K(x) =\frac{1}{\sqrt{2\pi}} e^{-\frac{x^2}{2}}$. The smoothing bandwidth $b$ was set to $1.36$ months for the estimation of the mean and to $4.59$ months for the estimation of the smoothed covariance. Estimation of cross-covariances used a $2.0$ month bandwidth for all the $X(t)$ considered. These bandwidths were obtained by 10-fold cross-validation (CV) \citep{Yao05, Yang2011}. In particular, to prevent under-smoothing we use the largest bandwidth possible such that the associated mean error is within one standard error of the minimum. 
Computing the auto- and cross- covariance functions $C_X$ and $C_{X_1,X_2}$ is also based on data aggregation. We use a locally weighted least squares bi-linear smoother, $S_{2D}$, in order to fit a smooth surface. The exact procedure is described in the Appendix Sect. \ref{app1} and \ref{app2}.   
The estimation procedures for both $S_{1D}$ and $S_{2D}$  are carefully outlined along with their corresponding asymptotic behaviours in \cite{Yao05}.
For the purposes of applying FPCA on the PROBIT dataset we will denote the head circumference trajectory of the $i$-th subject as $X_i^{H}$ and the body length trajectory of the $i$-th subject as $X_i^{L}$. 

\begin{figure}[!ht]
\centering
    \includegraphics[width=0.985\textwidth]{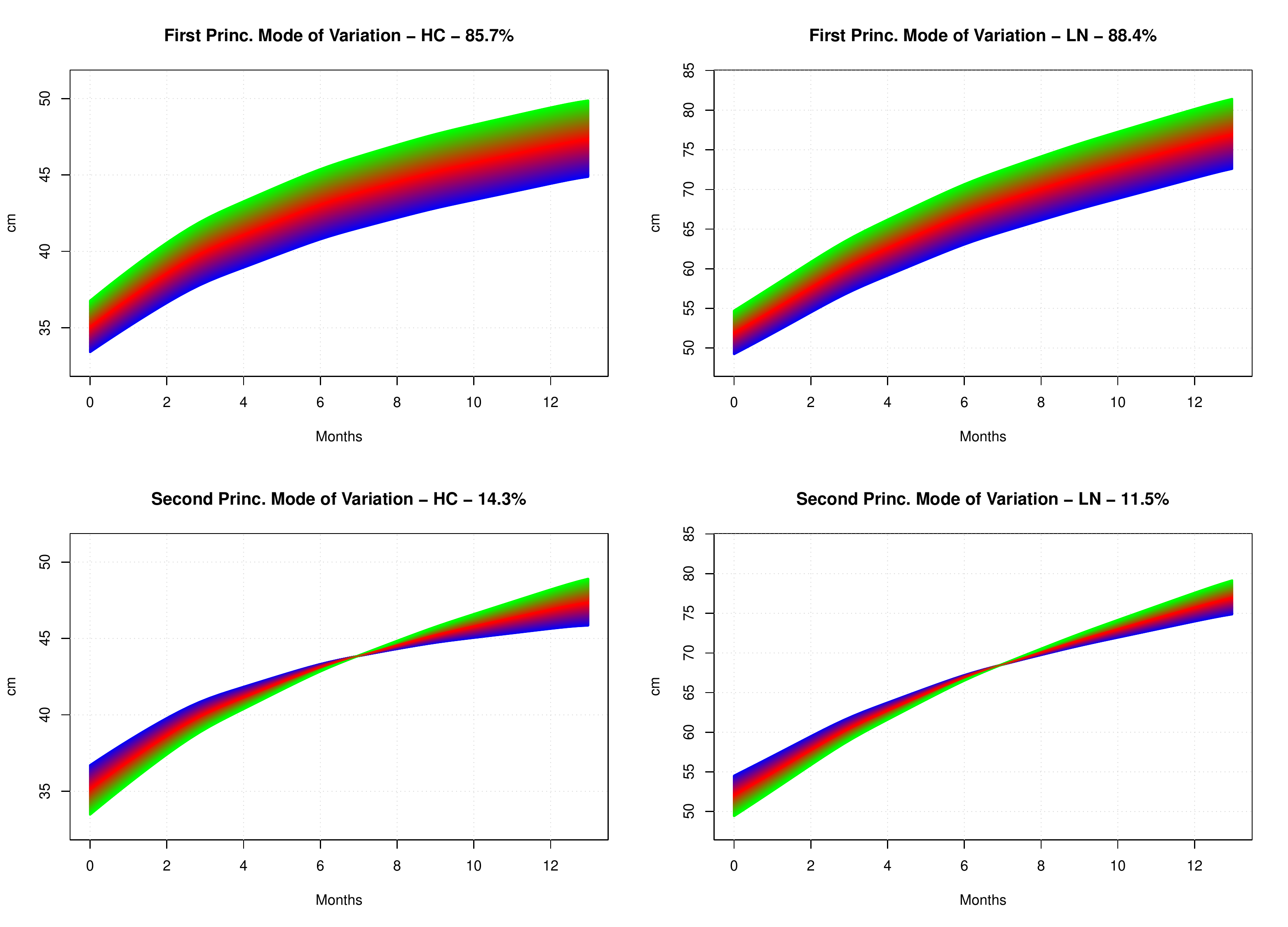} 
    \caption{
    The first two principal modes of variation for head circumference { (left panels, HC)} and body length {(right panels, LN)}. 
    Mean function $\mu_X(t)$ (red line), upper two standard deviations bounds shown in green colour and lower bounds shown in blue.
    Top row: first principal mode of variation, bottom row: second principal model of variation.}\label{ModesPlot} 
\end{figure}

The mean trajectories for the two functional covariates increase monotonically over the first year of life (Fig. \ref{ModesPlot}, red lines).
Using the FPCA results we examine the relevant eigenfunctions $\phi$ for both head circumference and body length. For both measurements the first mode of variation appears to approximate the overall intercept of the sample, i.e. if a child's size was \textit{larger} or \textit{smaller} than average (Fig. \ref{ModesPlot}, top row). The second principal mode of variation appears to move the edges of the trajectories in a complementary way: if the start of the trajectory is ``high", it leads to ``lower" values at the trajectories' end (Fig. \ref{ModesPlot}, bottom row) and vice versa, therefore it reflects variation of the overall change. 

Childhood intelligence is strongly associated with parental socio-demographic characteristics \citep{smithers2013impact, nisbett2012intelligence}. With this in mind we use a linear-mixed-effect (LME) model to adjust for parental influence on IQ. Preliminary investigation has demonstrated that the influence of time-invariant covariates at birth (e.g. maternal age-at-birth) on the IQ at age 6.5  {is constant} through-out the first year of a child's life. 
We also use the sex and birth-weight of the child as additional explanatory variables to mitigate strong sex- or birth-weight-related effects. 
We can therefore marginalise out the influence of the baseline covariates $Z$ on IQ by analysing the residuals of the following linear mixed-effects model: 
\begin{align}
 E\{IQ_{full}| Z \} =   &\beta_0 +  \beta_1 Z_{PE} + \beta_2  Z_{ME} + \beta_3  Z_{PAB} + \beta_4  Z_{MAB} + \nonumber \\  &\beta_5  Z_{MS} + \beta_6 Z_{EBF} + \beta_7 Z_{Sex} +  \beta_8 Z_{BW} + \gamma Z_{Hospital.} \label{LMEMReg}
\end{align}
The baselines for the categorical variables used were: \textit{less than secondary education} completed for $ME$ and $PE$, \textit{up to three months ($\leq 3$) of exclusive breastfeeding} for $EBF$ and \textit{no smoking} during pregnancy for $MS$. 

The residuals from this model,  ${IQ_{full}} -  \hat{IQ}_{full}$, { where $\hat{IQ}_{full}$ is the fitted value of $E\{ IQ_{full} | Z\}$ in the linear mixed-effects model \eqref{LMEMReg} given a covariate $Z$}, are treated as the new variable of interest, ${IQ_{res}}$, ``residual IQ"; from here on in referred to simply as IQ. A general summary of this model is available in the Appendix (Sect. \ref{TableSection}, \ref{LMEmodelBetas}). We note that, while this rather extensive model explains a substantial portion of the  overall IQ performance (conditional fit measuring explained variation in linear mixed effects models $\Omega^2_0: 0.379$ \citep{xu2003measuring}), it does not allow for time-varying estimates regarding the significance of different predictors. Since the PROBIT study is a cluster-randomised trial with hospitals and affiliated polyclinics as units of clustering, we control for this through the inclusion of the random effect $Z_{Hospital}$. 

Complementary to the original growth curves we also use their normalised version. The normalisation procedure is based on the FPCA using the original curves as input, {and as such} is an in-sample normalisation. For each variable $X(t)$ we have estimates for the mean $\mu_X(t)$ as well as the standard deviation $\sqrt{C(t,t)}$ at each time-point $t$ from Eq. \ref{AutoCovOper}. Using these the normalised sample can be directly estimated. One first subtracts the $\mu_X(t_j)$ from the relevant measurements and then divides by $\sqrt{C(t_j,t_j)}$. Chiou et al.  present this normalisation in more detail in \cite{chiou2014multivariate}. 
Using normalisation through FPCA is beneficial because PROBIT  children are consistently above the WHO standards for head circumference, body length and body weight due to the inclusion criteria of term-births with birth-weight $\geq 2.5 kg$. Normalising based on the WHO standards would therefore be possible but the resulting trajectories would not constitute an appropriately normalised sample (for example the mean trajectory of any of the longitudinal variables $X(t)$ would be consistently above zero).


\begin{table}[!ht]
\begin{center}
\caption{Cumulative  fraction of variance explained (FVE)  associated with the first $k$ functional principal components for each growth variable examined. (Rounded to two significant digits.)}\label{FVEtable} 
\begin{threeparttable}
\begin{tabular}{lccc}
\headrow
$X(t)$	&	 $k=1$	&	$k=2$	&	$k=3$ \\
Head Circum. & 85.65 & 99.98 & 99.99 \\  
Body Length  & 88.43 & 99.92  & 99.99  \\  
\hline  
\end{tabular}

\end{threeparttable}
\end{center}
\end{table}

\section{Associating Growth Trajectories and IQ}

\begin{figure}[!hb]
\centering
    \includegraphics[width=0.95\textwidth]{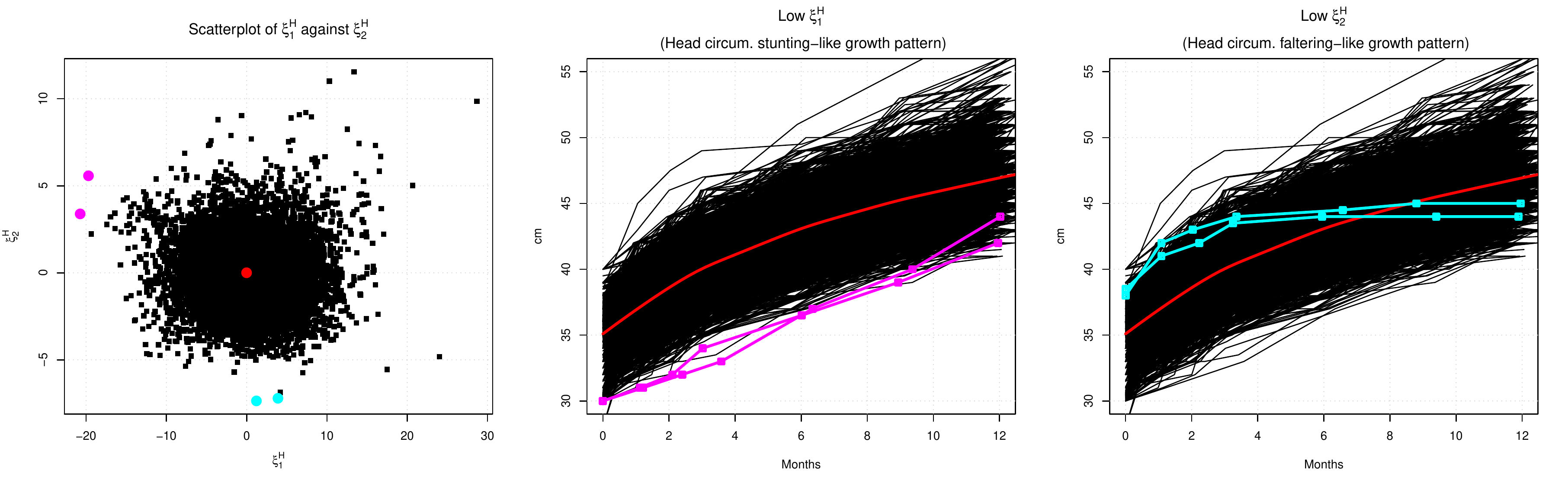} 
    \caption{The scatter-plot of functional principal components scores ({\color{black}leftmost} plot) allows easy identification of children with unusual growth patterns by looking at points that might diverge from the population norm. For example, the magenta points appear low in terms of $\xi_1$. Examining them shows that indeed these children appear significantly lower in terms of overall head circumference as their cohort counterparts. Similarly, the cyan points that are low in terms of $\xi_2$ appear to stagnate in terms of head circumference growth and, while appearing large during their first months of life, they subsequently drift to below-average values by the time they reach 12 months old. This pattern can be described as growth faltering, while the pattern seen in the magenta outliers is akin to growth stunting. The red-line corresponds to the functional mean.}\label{UnusualKids} 
\end{figure}

As seen in  Fig. \ref{ModesPlot} and \ref{UnusualKids}, the two principal modes of variation encapsulate easily interpretable physical trajectory patttens. 
The relevant fraction-of-variance-explained (FVE) measurements (Table \ref{FVEtable}) also suggest that two modes of variation are adequate. In both cases at least 95\% of the overall variance is retained by the first two FPCs. This further emphasises the parsimonious description that can be derived using FPCA; a significant proportion of the variation around the mean trajectories $\mu_x(t)$ can be quantified in a very low-dimensional domain (here with only two dimensions). We note that this proportion corresponds to denoised data variation and not the variation associated with the potentially noisy original measurements. 

Through FPCA we have the ability to directly estimate the levels of longitudinal trajectories at age 12 months for the physical attributes examined by generating imputed estimates of the trajectories. That is important because not all children are measured at the same times; there is some inherit variability of the measurement times. For example, within PROBIT the timing of the sixth visit was on average at 12.1 months of age. Nevertheless the standard deviation of these timings was 0.21 months, showing that children could commonly have nearly half a month age difference when they are measured. 
The variability in the timings of the visits provides a challenge to standard longitudinal data analysis approaches. Random-effects approaches like the Berkey-Reed growth curve model \citep{berkey1987model} are applicable but strongly parametric. Parametric approaches utilise a predefined number of particular variation patterns (e.g. differences in mean and deceleration) to model the growth dynamics and are less flexible than FDA-based models \citep{zhang2012nonparametric,chiou2016pairwise}. Through FDA-based models we can have sample-specific growth patterns that can encapsulate variability more efficiently. 
Using the imputed trajectories we can get imputed trajectories in time, allowing us to pick an arbitrary time point at which to compare curves. Our procedure provides denoising and smoothing. We found that, for children included in PROBIT 1, head circumference increased from $35.1 \pm 1.0 \text{cm}$ at birth to $46.9 \pm 1.4 \text{cm}$ by the time a child was 12 months old. This is consistent with the cross-sectional means of $35.0 \pm 1.5$ and $47 \pm 1.6 \text{cm}$ for the birth and 12-months head circumference respectively. Further details on imputation are given in the Appendix Sect. \ref{app1}.
In terms of body length, children increased from $52.0 \pm 1.8 \text{cm}$ at birth to $75.8 \pm 2.4 \text{cm}$ by the time a child was 12 months old. 
In addition, FPCA allows us to immediately identify children with unusual growth patterns (Fig. \ref{UnusualKids}) through the examination of their FPCA scores. 
This demonstrates further the benefits of using an FPCA approach for longitudinal growth data. 
	
\subsection{Linear regression approach}

We investigate the relation between head circumference growth during infancy and IQ in childhood by using the associated  FPCA scores within the context of a linear regression analysis. 
As mentioned, the first mode of variation of head circumference FPCA results (Fig. \ref{ModesPlot}, top left) reflects the difference between smaller and large overall sizes. Similarly, the  second mode of variation of head circumference FPCA results  (Fig. \ref{ModesPlot}, bottom left) reflects the difference between lower and higher rates of growth than average. 
This important insight should not be over-interpreted. For example the correlation between birth head circumference and $\xi_1^H$ is $\sim 0.40$. As expected, a child's overall head-size during infancy cannot be fully characterised at birth; the $6$-month and $9$-month visits have much stronger correlations ($0.85+$). Similarly, regarding the ``overall speed'' of growth, the correlation between $\xi_2^H$ and the difference between the $12$-month measurement and the birth head circumference is $\sim 0.75$, while it attains highest correlation ($0.85+$) when looking at the difference between the $12$-month measurement and the 2-month head circumference. In short, the first mode captures the overall size and the second mode the rate of growth.
 
We use the variables $\xi_1^{H}$ and $\xi_2^{H}$ as quantitative surrogates of a child's head circumference growth trajectories. $\xi_1^{H}$ and $\xi_2^{H}$ are constructed by using the estimated ${\phi}_1^H$ and  ${\phi}_2^H$ respectively, $\phi_1^H$ and $\phi_2^H$ serve as the axes onto which we project our data (i.e. contrary to a standard axes system where the axis is defined by unit-vectors, we use unit-functions - the FPCs $\phi$). The variables are normalised to have unit-variance; by construction $E\{ \xi_i \} = 0$. 
In addition, we also control for the interaction between the two modes. 
The actual model we use is as follows:
\begin{align}
E\{IQ_{res}| \Xi \} =  \beta_0 +  \beta_1^{H} \xi_1^{H}  + \beta_2^{H} \xi_2^{H} + \beta_3^{H} \xi_1^{H} \xi_2^{H}. \label{LM1}
\end{align}

\begin{figure}[!ht]
\centering
\includegraphics[width=0.99\textwidth]{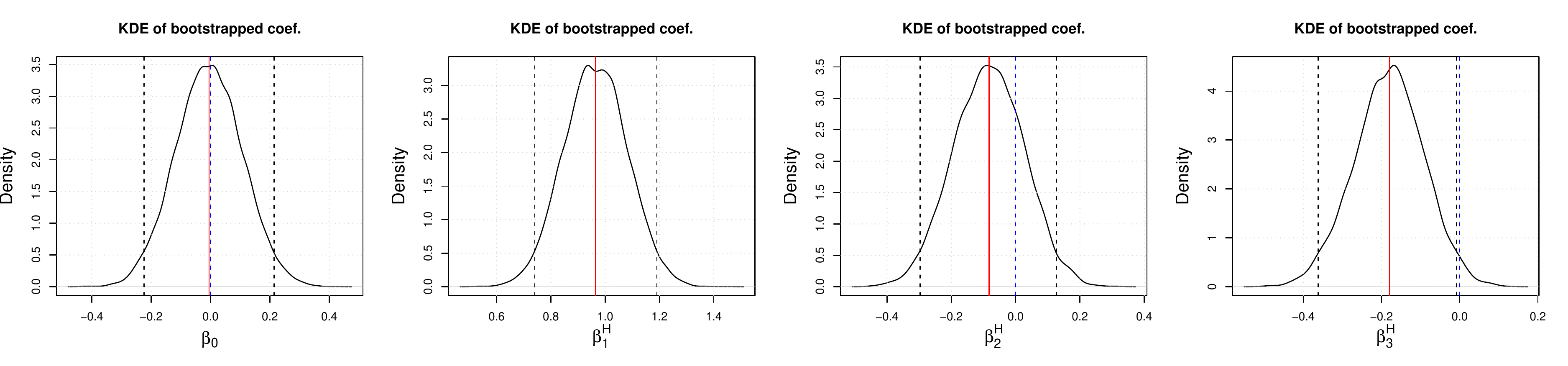}
    \caption{Bootstrap-generated {kernel density estimators (KDEs)} of the estimated $\beta$ of the model in Eq. \ref{LM1}. The red line shows the estimated $\beta$ from the data and the black dashed lines show the $95\%$ confidence intervals; the blue line denotes zero where relevant. The second plot from the left shows that the overall head circumference size has a statistically significant effect in the determination of later age IQ even when accounting for height information. $p$-values for $\beta_{0 \dots 3} = \{0.96, <0.001, 0.45, 0.05\} $.} \label{LM1plot} 
\end{figure}

As shown in Fig. \ref{LM1plot}, the estimated $\beta$'s suggest that the overall head circumference size during infancy ($\beta_1^H$) is significantly associated with higher IQ: 1 SD in terms of $\xi_1^{H}$  translates to one additional IQ point. The 95\% confidence intervals for this estimate and for estimates presented below are generated using 10000 bootstrapped samples, unless otherwise stated. On the other hand, the rate of growth is not significantly associated with the residual IQ outcome during the first year of life. The slightly negative (and statistically insignificant) effect that a higher growth rate conveys is due to higher growth rates being weakly associated with smaller head sizes, which are in turn associated with lower IQ; this is potentially a regression-to-the-mean phenomenon. The interaction between the two modes of variation appears marginally statistically significant in its association with IQ. Overall, children in PROBIT with overall large head circumferences during their first year of life have consistently higher IQ scores at age 6.5 years. On the contrary, large head circumference growth rate during the first year of life did not appear to be significantly associated with higher IQ scores at 6.5 years of age. Exact values for the coefficient estimates are reported in Table \ref{tableLM1}.

We also assess whether the prediction of IQ from head circumference measurements is improved by adding body length as the second predictor.
We perform the same analysis as with the model of Eq. \ref{LM1} but this time we also add the information about body length growth. The simultaneous modelling of $IQ_{res}$ using both the head circumference as well as the body length information controls for growth patterns that are better reflected in body length instead of head circumference.

\begin{figure}[!t]
\centering
    \includegraphics[width=0.995\textwidth]{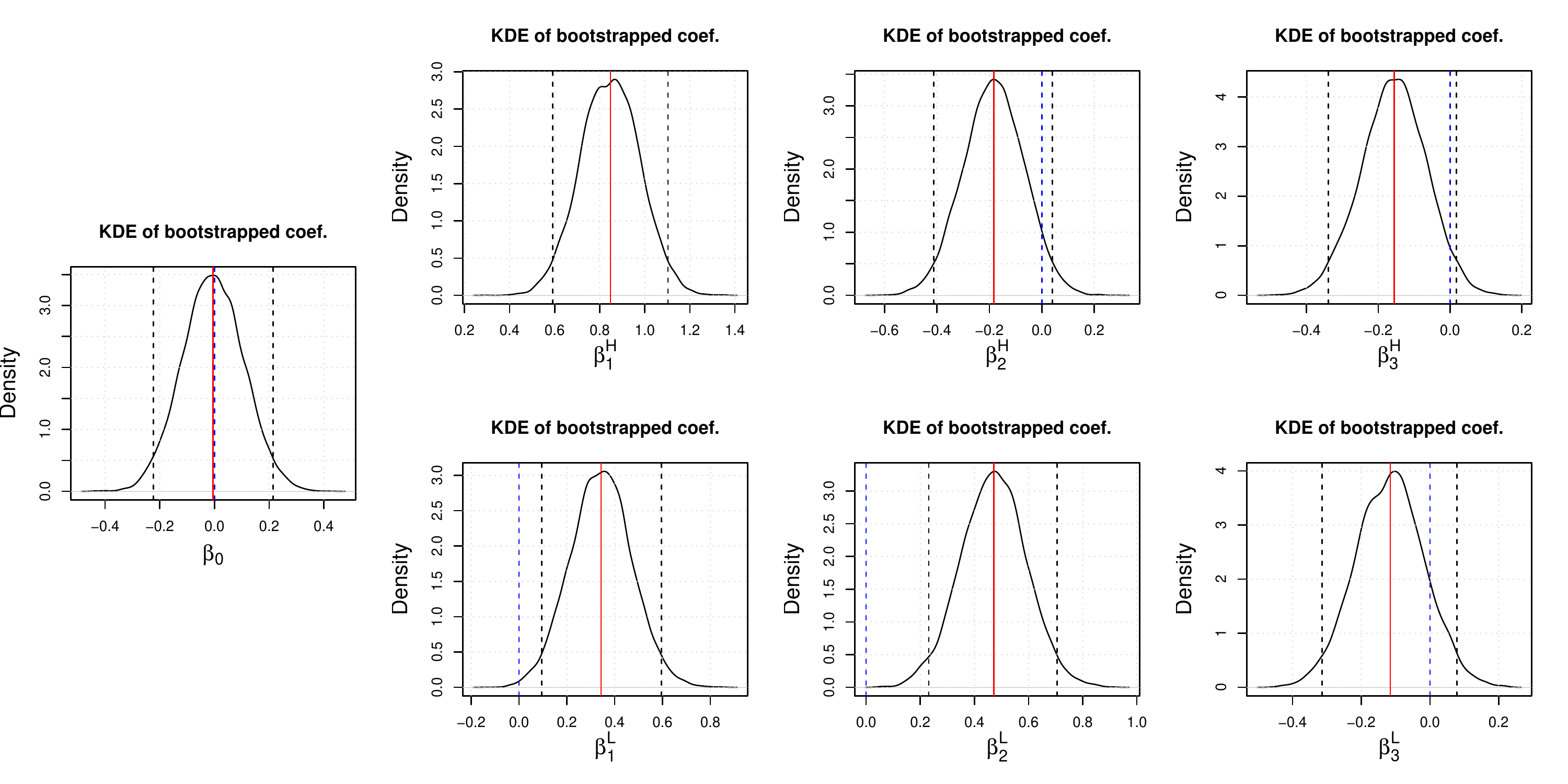} 
    \caption{Bootstrap-generated { kernel density estimators (KDEs)} of the estimated $\beta$  of the model in Eq. \ref{LM2}. The red line shows the estimated $\beta$ from the data and the black dotted lines show the $95\%$ confidence intervals; the blue line denotes zero where relevant. The overall head circumference size has a statistically significant effect in the determination of later age IQ. The effect of the growth rate of the head circumference from birth to 12 months has little statistical significance but the growth rate of the body length does appear statistically significant suggesting different interpretation of the growth dynamics between head circumference and body length. The $p$-values for $\beta_{0, {1^H \dots 3}^H,  {1^L \dots 3}^L}$ are $\{0.95, <0.001, 0.12, 0.10, 0.01, <0.001,  0.25\} $ respectively.}\label{LM2plot} 
\end{figure}

Body length information is once again assumed to be quantified by FPC scores of the body length. We denote the scores associated with the first and second principal mode of variation in terms of body length as $\xi_1^{L}$ and $\xi_2^{L}$ respectively. As before, the interaction between the scores $\xi_1^{L}$ and $\xi_2^{L}$ is also added in the model:
\begin{align}
E\{IQ_{res}| \Xi \} =  \beta_0 +  \beta_1^{H} \xi_1^{H} + \beta_2^{H} \xi_2^{H} + \beta_3^{H} \xi_1^{H} \xi_2^{H} + \beta_1^{L} \xi_1^{L}  + \beta_2^{L} \xi_2^{L} + \beta_3^{L} \xi_1^{L} \xi_2^{L}. \label{LM2}
\end{align}
This model incorporates information from both growth processes concurrently. The bootstrapped KDEs for the distribution of the coefficients are shown in Fig. \ref{LM2plot}; the confidence intervals are reported in Table \ref{tableLM2}. Again, a large overall head circumference appears to have the strongest effect. Interestingly, the second strongest effect is the body length growth rate. That means that aside from a child's ``head size'' status, the rate over which a child's body length grows is significantly associated with their future IQ performance. The growth rate of the head circumference is only marginally statistically significant in its association with IQ at 6.5; this was also the case with the overall length of the infant. We thus find that the growth pattern characterised by the head circumference's first principal mode of variation portends statistically significant differences in IQ performance. Notably, the effect of the rate of growth in body length was also statistically significant, suggesting that growth faltering in terms of body length might have more severe impact on later IQ than faltering in terms of head circumference.

\subsection{Correlation analysis approach and the functional concurrent model}

Following the original investigation of particular growth patterns using  linear models, we study the correlations between the estimated scores $\Xi$ and $IQ_{res}$. This allows us to assess directly how a particular growth pattern is correlated with $IQ_{res}$ and whether certain growth trajectories are associated with lower later-age IQ scores. 

The results regarding the correlation of $\xi_1$ of $X(t)$ and $IQ_{res}$ are shown in Table \ref{EaseCorrs}.
We see that the correlation between the overall head circumference growth trend and IQ is stronger than the one with body length. For the second component $\xi_2$, the correlations are substantially weaker, in line with our 
 earlier findings using a linear regression approach (Fig. \ref{LM2plot}). 
These preliminary results suggest that the major modes of variation, both in terms of head circumference and body length, have weak but still statistically significant associations with later age IQ.  {When using raw differences between $X(t=\text{birth})$ and $X(t= \text{12 months})$, the results were marginally positive but again very weak ($0.028$ for head circumference and $0.039$ for body length).} The statistically insignificant negative correlation between the second slope-like mode of variation and $IQ_{res}$ can be potentially explained by the fact that most children with positive ``slopes'' have small body length/head circumference, which might attenuate the positive influence of growth. Similarly negative ``slopes'' are mostly prominent for children who are big overall.

\begin{figure}[!hb]
\centering
    \includegraphics[width=0.99\textwidth]{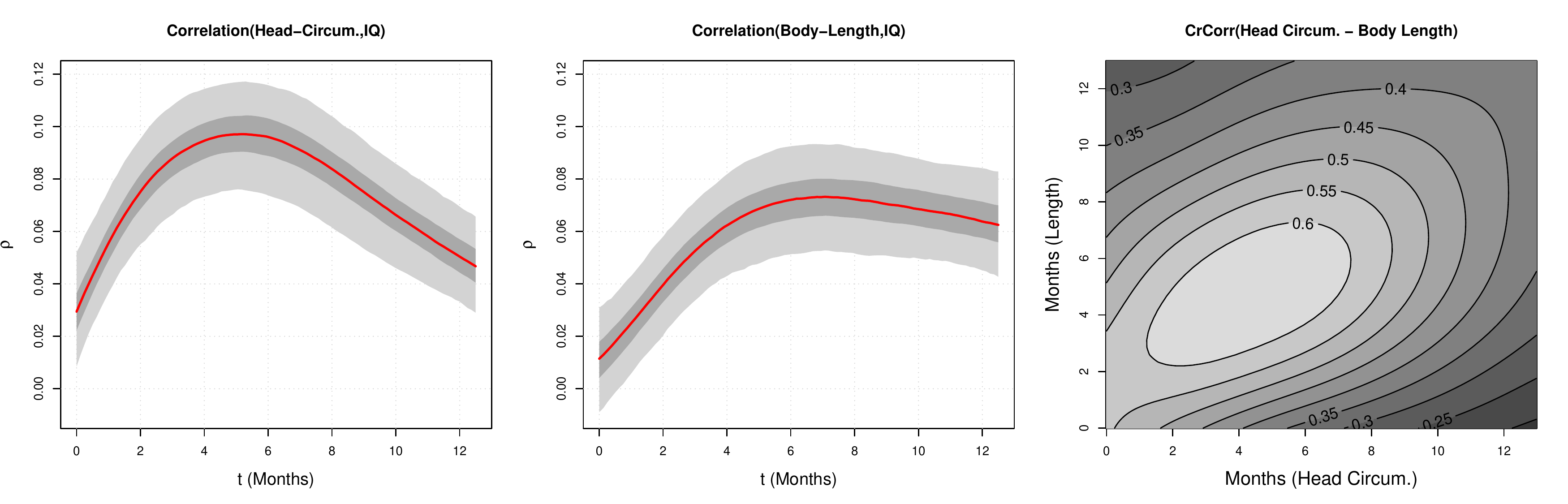} 
    \caption{Functional correlation estimates between residual IQ and each of the longitudinal variables $X(t)$ examined (left and middle plots). The red line shows the estimated correlation function $\rho(t)$ from the data; the light and dark gray bands are the $95\%$ and $50\%$ confidence intervals respectively. The cross-correlation between the two functional variables is shown in the right plot.}\label{FuncCorrs} 
\end{figure}

To obtain estimates for correlation between a scalar $Z$ (here, $IQ_{res}$) and functional variable $X(t)$, we adopt the singular functional correlation procedure of 
\cite{Yang2011}. The resulting correlation functions can be seen in Fig. \ref{FuncCorrs} (left-hand side and middle plot). The pointwise confidence intervals were generated using nonparametric bootstrap by resampling subjects ($k_{boot} = 1000$). Overall, the correlation trajectory for head circumference is higher than the correlation trajectory for body weight. Additionally it appears that the correlations are non-zero over almost the entire domain of the first year after birth. 
We found that 
while head circumference at $\sim 5$ months shows the highest correlation with later age IQ, this association decreases for older ages. Body length on the other hand is not significantly correlated with IQ on early ages but then retains a stable correlation level after approximately 4 months of age (Fig. \ref{FuncCorrs}, middle plot). As expected, the correlations are low in absolute terms but are significant.

\begin{table}[!hb]
\begin{center}
\caption{Pearson correlation between the first two functional principal components (FPCs), $\xi_1^{L/H}$ and $\xi_2^{L/H}$, and $IQ_{res}$, respectively. ($95\%$ confidence intervals in parenthesis generated by bootstrapping.)} \label{EaseCorrs} 
\begin{threeparttable}
\begin{tabular}{lcc}
\headrow
$X(t)$ &  $\rho(FPC_1, IQ_{res})$  & $\rho(FPC_2, IQ_{res})$ \\
Head Circum. &  0.07 (0.06, 0.09) & -0.01 (-0.03, 0.01)  \\ 
Body Length & 0.06 (0.04, 0.08) &   0.03 (0.01, 0.04)  \\   
\hline  
\end{tabular}

\end{threeparttable}
\end{center}
\end{table}

We also investigate the correlation between the head circumference growth and body length across time. This correlation is quantified as a surface rather than a simple one dimensional function as it is defined across time-scales; head circumference at age $s$ and body length at age $t$. The estimated correlation surface in $s$ and $t$ is visualised in the right-hand side plot of Fig. \ref{FuncCorrs}. As expected a child's head circumference during infancy and body length are substantially correlated  with each other throughout the first 12 months. The correlation remains above a $0.30$ level throughout infancy.  We observe 
a decline in that correlation as children grow: early infancy head circumference measurements are more strongly correlated with later infancy body length measurements than early infancy body length measurements are with later-infant-life head circumference measurements. 
This decreasing correlation between head circumference and body length with age during infancy is consistent with other studies \citep{geraedts2011association}. Notably through this methodology we are able to directly assess the correlation at two different ages (e.g. head-circumference at 4 months of age against body length at 3 or 11 months of age) in a natural and continuous way.



\begin{figure}[!ht]
\centering
    \includegraphics[width=0.99\textwidth]{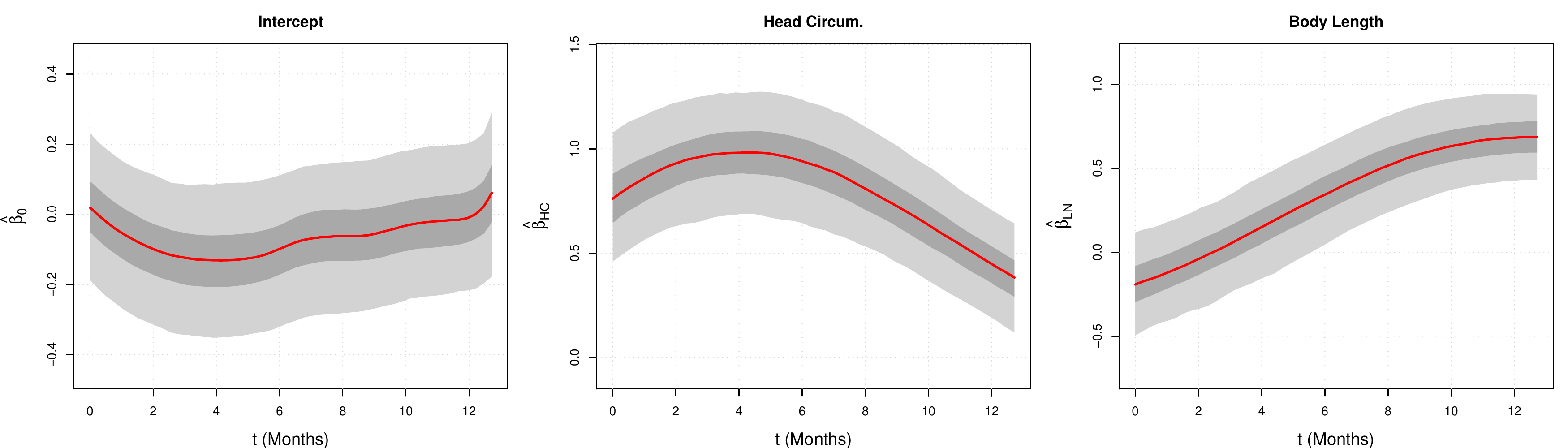} 
    \caption{Functional $\beta$ estimates between residual IQ at 6,5 years and each of the longitudinal variables $X(t)$ examined in Eq.\eqref{FCR1} obtained from the functional concurrent regression (FCR). The red line shows the estimated $\beta$ from the data, the dark gray band is the $50\%$ and the light gray band the $95\%$ confidence interval. }\label{FCRbetas} 
\end{figure}

We complement the correlation analysis by a functional concurrent regression (FCR) model \citep{csenturk2010functional, csenturk2011varying}. 
This model pinpoints the changing nature of the influence of growth characteristics on subsequent IQ, as it provides the relationship for each fixed age, over a large domain of ages.  
In the current longitudinal design FCR can extract valuable information regarding the statistical significance of infant growth variables for subsequent IQ, as these variables evolve over time, where
\begin{align}
E\{IQ_{res}| X^{H}(t), X^{L}(t) \} =  \beta_0(t) +  \beta^{H}(t)   X^{H}(t) +  \beta^{L}(t)   X^{L}(t). \label{FCR1}
\end{align}
This model features functional regression coefficients for both head circumference, $ \beta^{H}(t)$, as well as body length, $\beta^{L}(t)$. An intercept term, $\beta_0(t)$, is also included, and $X^H(t)$ and $X^L(t)$ are head circumference and body length at age $t$, respectively. We do not include baseline covariates $Z$ (Table \ref{TableVars}) because they appear to have a constant effect during infancy, based on preliminary investigation (results not shown). 
The resulting $\beta(t)$ (Fig. \ref{FCRbetas}; $k_{\text{boot}} =1000$) indicate the time periods that are more predictive for subsequent IQ. Head circumference is most influential between 3 and 5 months. The influence of body length becomes increasingly stronger with age: at birth it is statistically insignificant while by 12 months of age it is almost as significant as head circumference. As expected, the overall intercept is close to 0. 
As a general trend, head circumference is overall more informative than body length but the influence of the two measures becomes largely similar by 12 months. 
Overall, Figure \ref{FCRbetas} shows that the influence of body length growth consistently increases during infancy, while the association of head-circumference growth with IQ is larger in early months. 
These results are in accordance with the correlation analysis. 
The different shapes between the correlation trajectory of body length and IQ (Fig. \ref{FuncCorrs}, middle plot) against that of $\beta^{LN}(t)$ (Fig. \ref{FCRbetas}, right-hand side plot) can be attributed to the fact that, while the correlation analysis did not incorporate head circumference, head circumference is included in the FCR model. The advantage of FCR is that it includes the variance-covariance matrix of the longitudinal covariates and accounts for common variance patterns. 

\section{Discussion}

We demonstrate that methods from functional data analysis, notably functional principal component analysis, enable researchers to study the relationships between several longitudinal growth curves that are usually obtained from intermittent measurements and later outcomes. 
The proposed eigenanalysis indicates an association 
of early childhood IQ with head circumference growth during infancy, supporting previous findings \citep{fattal2009growth}. 
Using the proposed approach, we find 
that changes of 1 SD in terms of overall head circumference size during infancy are associated with changes by approximately 1 IQ point. On the other hand, our findings suggest that changes in body length growth patterns are not associated with IQ differences, given head circumference information.  
 In addition, the functional concurrent model indicates that early infancy body length measurements are less informative than head circumference measurements for neurocognitive development and the focus of developmental research should be primarily on the recordings of head circumference if one aims to quantify associations between growth patterns and early childhood intelligence. Our findings also suggest that the correlation and the linear association of head circumference measurements with IQ peak between 3 and 5 months of infancy. 

Correlation plots are highly informative in applications to growth studies. Employing function-to-scalar correlation we are able to identify periods of increased correlation. Through function-to-function correlation plots one can quantify time-varying correlations of head circumference and body length during the infancy of a child.  Finally, functional concurrent models enables us to investigate the age-varying significance of infancy growth in relation to later age IQ. 


While the growth curves are typically sampled only at a few randomly varying discrete time-points, functional models make it possible to overcome this limitation. Functional models treat these discrete growth readings as continuous random processes that can be adequately represented by functional principal components. One may then highlight the main modes of variation that quantify growth stunting or increased overall growth through the first component, and growth faltering or acceleration through the second component.  

The focus of this work is the association of growth data and early childhood cognitive performance under a fully non-parametric framework. Nevertheless the methodology presented can be used for other applications where irregularly measured covariates relate to a scalar outcome. The future goals of this project are therefore two-fold. First, the proposed methodology can be used to make meaningful inference from other developmental datasets, thus allowing the creation of more data-driven growth charts as well as the detection of potentially problematic growth trajectories.  Second, by taking advantage of the surrogate variables generated (FPC scores, covariance surfaces, etc.), it may be possible to  infer associations between populations of children that share common developmental patterns. \cite{pigoli2018statistical} have already presented such an application on the Phylogenetics of Romance languages which utilises the associated covariance functions. Further extensions of using surrogate functional variables open the avenue of examining growth curves under a mixed framework where amplitude- and time-variation are examined separately \citep{hadjipantelis2015unifying}; in such a model the warping functions of the sample would encapsulate growth-timing information.

\section*{acknowledgements}
This study was supported by the Bill \& Melinda Gates Foundation (OPP1119700). The article contents are the sole responsibility of the authors and may not necessarily represent the official views of the Bill \& Melinda Gates Foundation or other agencies that may have supported the primary data studies used in the present study.





\bibliography{fpReportBib}



\appendix
\section{Appendix}

\subsection{Functional principal component analysis for sparse longitudinal data} \label{app1}
We introduce functional principal component analysis (FPCA) for longitudinal data through conditional expectation, called the PACE method \citep{Yao05,YMW2005b}. Let $X$ be a smooth random function with mean $\mu(t) = EX(t) $ and covariance $C(s,t) = \text{Cov}(X(s), X(t))$, where $X$ is square integrable on a compact time interval $\mathcal{T}$. Let $X_i$, $1 \leq i \leq n$, be independent realisations of $X$ which correspond to random trajectories for the $i$-th subject, respectively. According to the Karhunen-Lo\`{e}ve theorem \citep{K1946,L1946}, $X_i$ admit the following representation $X_i(t) = \mu(t) + \sum_{k=1}^\infty \xi_{i,k} \phi_k(t)$,
where $\phi_k$ are orthonormal in $L^2(\mathcal{T})$ such that $C(s,t) = \sum_{k=1}^\infty \lambda_k \phi_k(s) \phi_k(t)$, $s,t \in \mathcal{T}$, with nonincreasing and positive eigenvalues $\lambda_1 \geq \lambda_2 \geq \cdots > 0$, and $\xi_{i,k}$, $k \geq 1$, are uncorrelated random variables with mean 0 and variance $\text{Var}(\xi_{i,k}) = \lambda_k$. 

In conventional FPCA, we estimate mean $\mu$ and covariance $C$ cross-sectionally based on dense observations $X_i$ over time domain $\mathcal{T}$, and the pairs of eigenvalues and eigenfunctions $(\lambda_k, \phi_k)$ are obtained by a spectral decomposition of estimated covariance. For more details, refer to \cite{horvath2012inference, wang2016functional}. In many longitudinal studies, however, sparse and noisy observations are collected from underlying random trajectories. Let $T_{ij}$ and $\epsilon_{ij}$ be randomly sampled observation time points and measurement errors at observations, respectively, and we have longitudinal observations $Y_{ij} = X_i(T_{ij}) + \epsilon_{ij}$, $1 \leq j \leq N_i$, for the $i$-th subject. The measurement errors are typically assumed to be iid with mean 0 and variance $\text{Var}(\epsilon) = \sigma_\epsilon^2$, and independent of a random trajectory $X_i$ and observation time $T_{ij}$. Also, the number of longitudinal observations $N_i$ are random integers independent of all other random quantities. We note that $E(Y_{ij}|T_{ij}=t)= \mu(t)$ and $\text{Cov}(Y_{ij}, Y_{i\ell} | T_{ij}=s, T_{i\ell}=t) = C(s, t) + \sigma_\epsilon^2 \cdot \mathbb{I}(s=t)$. 

Local smoothing estimation \citep{FG1996} for mean $\mu$ and covariance $C$ can be applied for one- and two-dimensional smooth function and surface, respectively. First, the mean function can be estimated by a locally linear smoothing estimator $\hat{\mu}(t)$ define by $\hat{\beta}_0$, where $(\hat{\beta}_0, \hat{\beta}_1)$ are pairs of the minimizers for a weighted least squares criterion 
\begin{align}
	\sum_{i=1}^{n} \sum_{j=1}^{N_i} K_1\left( \frac{T_{ij} -t}{h_0} \right) \{ Y_{ij} - \beta_0 - \beta_1 (t-T_{ij}) \}^2 \nonumber
\end{align}
for each $t \in \mathcal{T}$ with respect to $(\beta_0, \beta_1) \in \mathbb{R}^2$, where $K_1$ is a compactly supported univariate kernel function and $h_0$ is a positive bandwidth. Then, the covariance function can be estimated by a locally linear smoothing estimator $\widehat{C}(s,t)$ defined by $\hat{\gamma}_0$, where $(\hat{\gamma}_0, \hat{\gamma}_{1}, \hat{\gamma}_{2})$ are tuples of the minimizers for a weighted least squares criterion
\begin{align}
	\sum_{i=1}^{n} \sum_{j \neq \ell}^{N_i} K_2\left( \frac{T_{ij} -s}{h_1}, \frac{T_{i\ell} -t}{h_2} \right) \big\{ {C}_i(T_{ij}, T_{i\ell}) - \gamma_0 - \gamma_{1}(s-T_{ij})- \gamma_{2}(t-T_{i\ell}) \big\}^2 \nonumber
\end{align}
for each $s, t \in \mathcal{T}$ with respect to $(\gamma_0, \gamma_{1}, \gamma_{2}) \in \mathbb{R}^3$. Here, ${C}_i(T_{ij},T_{i\ell}) = (Y_{ij} - \hat{\mu}(T_{ij}))(Y_{i\ell} - \hat{\mu}(T_{i\ell}))$ are raw covariances, the kernel $K_2$ is a compactly supported bivariate kernel function and $h_1, h_2$ are positive bandwidths. The variance of the measurement errors $\sigma_\epsilon^2$ is also a parameter of interest. Since the maximal values of the covariance surface should be retained on diagonal, a locally quadratic smooth estimation provides improved estimation \citep{YMCDFLBV2003,Yao05,YMW2005b}. 
For comprehensive reviews, see \cite{M2009,wang2016functional}.

Finally, eigenfunctions and eigenvalues are obtained by the successive solutions of the following equations
\begin{align}
	\int_\mathcal{T} \widehat{G}(s,t) \hat{\phi}_k(s)~ds = \hat{\lambda}_k \hat{\phi}_k(t) \quad (t \in \mathcal{T}), \nonumber
\end{align}
where $\int_\mathcal{T} \hat{\phi}_k(t)^2~dt = 1$ and $\int_\mathcal{T} \hat{\phi}_k(t) \hat{\phi}_\ell(t)~dt = 0$ for $\ell < k$. Discretisation is used to estimate the eigenfunctions \citep{RS1991, CM1997}.

{FPCA enables us to impute longitudinal outcomes. The individual FPC scores can be obtained from Eq. \eqref{compPACE}, say $\hat{\xi}_{i,k}$, with arguments being replaced by $\hat{\mu}$, $\widehat{C}$, $\hat{\phi}_k$ and $\hat{\lambda}_k$, respectively. borrowing information across subjects. We note that this PACE method provides the best prediction of $\xi_{i,k}$ under Gaussian assumptions and more generally provides the best linear predictors. It is robust to the violence of the Gaussian assumption \citep{Yao05, YMW2005b}. Functional variables $X_i$ can be reconstructed by the optimal $K$-dimensional linear approximation as in Eq. \eqref{CastroEq}, $\hat{X}_i(t) = \hat{\mu}(t) + \sum_{k=1}^K \hat{\xi}_{i,k} \hat{\phi}_k(t)$ for $t \in \mathcal{T}$. The reconstructed functional variables $\hat{X}_i$ provide imputed longitudinal outcomes.}

\subsection{Estimation of the cross-covariance function} \label{app2}

We compute the local smoothing estimators of the cross-covariances in Eq. \eqref{CrossCovOperReg}. First, we consider cross-covariances between a random function $X$ and a random variable $Z$. Let $(X_i, Z_i)$ be independent realisations of $(X,Z)$ with mean $(\mu_X, \mu_Z)$ and covariance $C_{X,Z}(t) = E(X(t) - \mu_X(t))(Z - \mu_Z)$, $t \in \mathcal{T}$. Similarly to the previous subsection \ref{app1}, $Y_{ij}$ are noisy and sparse observations for $X_i(T_{ij})$. We apply local smoothing $C_{X,Z}$. In particular, $\widehat{C}_{X,Z}(t)$ is defined by $\hat{\alpha}_0$, where $(\hat{\alpha}_0, \hat{\alpha}_1)$ are the minimizers of a weighted least squares,
\begin{align}
	\sum_{i=1}^{n} \sum_{j=1}^{N_i} K_1\left( \frac{T_{ij} -t}{h_0} \right) \big\{ {C}_i^{X,Z}(T_{ij}) - \alpha_0 - \alpha_{1}(t-T_{ij}) \big\}^2, \nonumber
\end{align}
for each $t \in \mathcal{T}$ with respect to $({\alpha}_0, {\alpha}_1) \in \mathbb{R}^2$. Here, $\hat{\mu}_X$ for $\mu_X$ is obtained by locally linear smoothing, $\hat{\mu}_Z$ for $\mu_Z$ by $\hat{\mu}_Z = n^{-1}\sum_{i=1}^n Z_i$, and ${C}_i^{X,Z}(T_{ij})  = (Y_{ij} - \hat{\mu}_X(T_{ij})) (Z_i - \hat{\mu}_Z)$ is a raw cross-covariance between $Y_{ij}$ and $Z_i$.

Next, we consider a cross-covariance between two random functions. Let $(X_{1i}, X_{2i})$ be independent realisations of $(X_1, X_2)$, a pair of two random trajectories with mean $(\mu_1, \mu_2)$ and covariance $C_{12}(s,t) = \text{Cov}(X_{1i}(s), X_{2i}(t))$, $s,t \in \mathcal{T}$. Similarly to the previous subsection \ref{app1}, $Y_{1,ij}$ and  $Y_{2,i\ell}$ are noisy and sparse observations for $X_{1i}(T_{ij})$ and $X_{2i}(T_{i\ell})$, respectively. We define $\widehat{C}_{12}(s,t)$ by $\hat{\eta}_0(s,t)$, where $(\hat{\eta}_0, \hat{\eta}_1, \hat{\eta}_2)$ are the minimizers of a weighted least squares
\begin{align}
	\sum_{i=1}^{n} \sum_{j \neq \ell}^{N_i} K_2\left( \frac{T_{ij} -s}{h_1}, \frac{T_{i\ell} -t}{h_2} \right) \big\{ {C}_{i}^{X_1,X_2}(T_{ij}, T_{i\ell}) - \eta_0 - \eta_{1}(s-T_{ij})- \eta_{2}(t-T_{i\ell}) \big\}^2, \nonumber
\end{align}
for each $s,t \in \mathcal{T}$ with respect to $({\eta}_0, {\eta}_1, {\eta}_2) \in \mathbb{R}^3$. Here, $\hat{\mu}_1$ and $\hat{\mu}_2$ are obtained by locally linear smoothing, and ${C}_{i}^{X_1,X_2}(T_{ij}, T_{i\ell}) = (Y_{1,ij} - \hat{\mu}_1(T_{ij})) (Y_{2,i\ell} - \hat{\mu}_2(T_{i\ell}))$ is a raw cross-covariance between $Y_{1,ij}$ and $Y_{2,i\ell}$ for each $i$-th subject. For details, see \cite{chiou2014multivariate, csenturk2011varying}.

\subsection{Design plots} \label{app3}
\begin{figure}[!ht]
\centering
\includegraphics[width=0.5\textwidth]{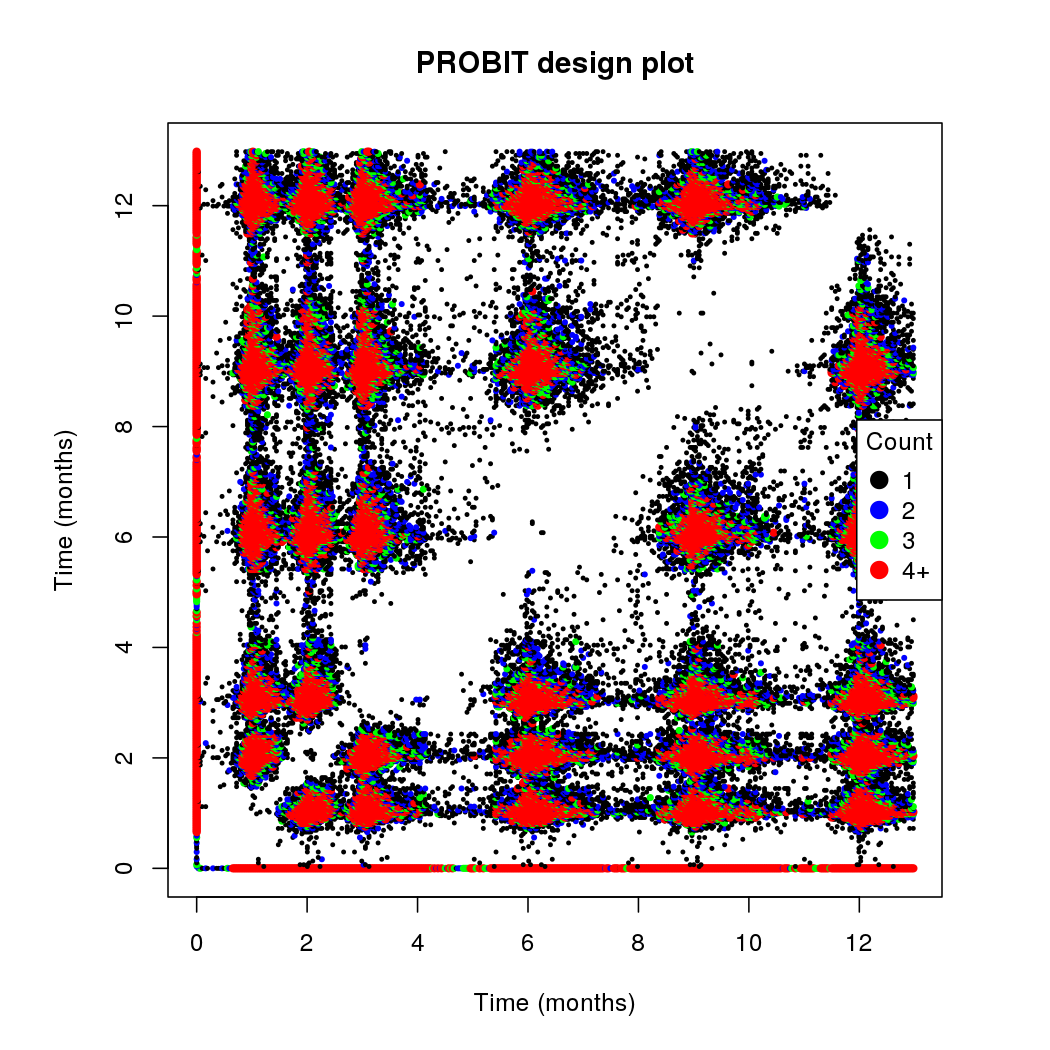} 
  \caption{The design plot showing the measurements times during the first-year follow-up. Expectedly the data have a strong clustering around the measurement times at two, three, six, nine and twelve months. The diagonal measurements are not shown. 
  }\label{DesignPlot} 
\end{figure}

We collect a pooled grid of observation times $\mathcal{T}_N = \{ t_1, \ldots, t_N \}$ such that $\mathcal{T}_N = \cup_{i=1}^n\cup_{j=1}^{N_i} \{T_{ij}\}$, where we have $N_i$ observation times $T_{i1}, \ldots, T_{iN_i}$ for each $i$-th subject. We define a $N$-by-$N$ binary matrix $U_i$ such that each $(j,k)$-component of $U_i$ has a value 1, if there exists a pair of observation time points $(T_{i\ell},T_{im})$ that equals $(t_j, t_k)$ for some $1 \leq \ell,m \leq N_i$, otherwise 0. Then, $U = \sum_{i=1}^n U_i$ represents total counts of observation times at the unique observation grid across all subjects with design clusters as given in Figure \ref{DesignPlot}.

\newpage
\subsection{Accompanying tables} \label{TableSection}

\begin{table}[!h]
\begin{center}
\caption{Estimated coefficient for the linear mixed effects model shown in Eq. \ref{LMEMReg} together with bootstrap confidence intervals. The strong influence of the \textit{Hospital} random effect is due to the spatial information encapsulated in it.  
} \label{LMEmodelBetas} 
\begin{threeparttable}
\begin{tabular}{lccc}
\headrow
Coefficient names & $\hat{\beta}_i$ & $2.5\%$ & $97.5 \%$ \\
$\beta_0$ (Intercept)      & 97.74  & 93.69 & 102.22 \\ 
$\beta_{1}$ (Common Sec.)  & 2.66   &  1.15& 4.22 \\ 
$\beta_{1}$ (Special Sec.) &  4.28  &  2.70& 5.87 \\ 
$\beta_{1}$ (Tertiary)     &  8.68  & 6.99& 10.32 \\ 
$\beta_{2}$ (Common Sec.)  &  2.56  & 1.30& 3.89 \\ 
$\beta_{2}$ (Special Sec.) &  6.85  & 5.61& 8.17 \\ 
$\beta_{2}$ (Tertiary)     & 11.78  & 10.26& 13.21 \\ 
$\beta_3$ (PAB)            & -0.18  & -0.25& -0.12 \\ 
$\beta_4$ (MAB)            & -0.04  &  -0.11& 0.02 \\ 
$\beta_5$ (Smoking=Yes)            & 0.20   &  -1.55& 1.92 \\ 
$\beta_{6}$ (3 to 6 M.)    & 0.75   & 0.12& 1.35 \\ 
$\beta_{6}$ (6+ M.)        & 0.20   & -1.01& 1.43 \\ 
$\beta_7$  (Male)          & 0.02   & -0.44& 0.48 \\ 
$\beta_8$ (BW)             & 1.43   & 0.84& 1.95 \\ 
$\gamma (\sigma_{\text{Hospital}})$  &  9.86 & 7.23& 12.42 \\ 
$\sigma_\epsilon$ (Residual) & 12.77 & 12.61 & 12.92 \\ 
\hline  
\end{tabular}

\end{threeparttable}
\end{center}
\end{table}

\begin{table}[!h]
\begin{center}
\caption{Estimated coefficient for the linear model shown in Eq. \ref{LM1} together with bootstrap confidence intervals. The strong statistical significance of the first principal component is clearly reflected. It suggests that the basic size categorisation (large vs. small) in a toddler's head-circumference is associated with the toddler's later age IQ.} \label{tableLM1} 
\begin{threeparttable}
\begin{tabular}{lcccc}
\headrow
Coefficient names & $\hat{\beta}_i$ & $2.5\%$ & $97.5 \%$  & $p$-values \\
$\beta_0$ (Intercept)      &  -0.01  & -0.22 & 0.21 & 0.96 \\ 
$\beta_{1}^H$ (FPC1)        &  0.96  &  0.74 & 1.19 & $<$0.001   \\ 
$\beta_{2}^H$ (FPC2)        & -0.09  & -0.30 & 0.13 & 0.45 \\ 
$\beta_3^H$   (FPC1 * FPC2) & -0.18  & -0.36 & -0.01 & 0.05 \\ 
\hline  
\end{tabular}

\end{threeparttable}
\end{center}
\end{table}

\begin{table}[!h]
\begin{center}
\caption{Estimated coefficient for the linear model shown in Eq. \ref{LM2} together with bootstrap confidence intervals. The strong statistical significance of the head circumference first principal component is clearly reflected as in model \ref{LM2}. Importantly, the acceleration categorisation (slowing vs. accelerating) in terms of body length is also strongly associated but with lower effect sizes.}\label{tableLM2}
\begin{threeparttable}
\begin{tabular}{lcccc}
\headrow
Coefficient names & $\hat{\beta}_i$ & $2.5\%$ & $97.5 \%$  & $p$-values \\
$\beta_0$ (Intercept)       &  -0.01  & -0.22 & 0.21 & 0.95 \\ 
$\beta_{1}^H$ (FPC1)        &   0.85  &  0.59 & 1.10 &  $<$0.001  \\ 
$\beta_{2}^H$ (FPC2)        &  -0.18  & -0.34 & 0.02 &  0.12 \\ 
$\beta_3^H$   (FPC1 * FPC2) &  -0.16  & -0.41 & 0.04 & 0.10 \\ 
$\beta_{1}^L$ (FPC1)        &   0.34  &  0.10 & 0.59 & 0.01   \\ 
$\beta_{2}^L$ (FPC2)        &   0.47  &  0.23 & 0.71 & $<$0.001 \\ 
$\beta_3^L$   (FPC1 * FPC2) &  -0.12  & -0.31 & 0.08 & 0.25 \\ 
\hline  
\end{tabular}

\end{threeparttable}
\end{center}
\end{table}

\end{document}